\documentclass[amssymb,prb,twocolumn,floats,amsmath,showpacs,superscriptaddress]{revtex4}
\usepackage[T1]{fontenc}
\usepackage{bm}
\usepackage{graphicx}
\usepackage{amssymb}
\usepackage{leftidx}
\usepackage{amsfonts}
\usepackage{amsmath} 
\usepackage[outdir=./]{epstopdf}
\usepackage{color}
\usepackage{float}

\begin{document}

\title{\large \textbf{Temperature dependence of photoluminescence lifetime
of
atomically thin WSe$_2$ layer}}

\def \FUW{Institute of Experimental Physics, Faculty of Physics, University
of Warsaw, ul. Pasteura 5, 02-093 Warsaw, Poland}

\def \LNCMI{Laboratoire National des Champs Magnetiques Intenses,
CNRS-UGA-UPS-INSA-EMFL, 25 rue des Martyrs, 38042 Grenoble, France}

\def \NHMFL{National High Magnetic Field Laboratory, Los Alamos, NM 87545, USA}
\def \ETH{Institute for Quantum Electronics, ETH Z\"{u}rich, CH-8093 Z\"{u}rich, Switzerland}

\author{A.~\L{}opion}\affiliation{\FUW}
\author{M.~Goryca}\affiliation{\FUW}\affiliation{\NHMFL}
\author{T.~Smole\'{n}ski}\affiliation{\FUW}\affiliation{\ETH}
\author{K.~Oreszczuk}\affiliation{\FUW}
\author{K.~Nogajewski}\affiliation{\FUW}\affiliation{\LNCMI}
\author{M.~R.~Molas}\affiliation{\FUW}\affiliation{\LNCMI}
\author{M.~Potemski}\affiliation{\FUW}\affiliation{\LNCMI}
\author{P.~Kossacki}\affiliation{\FUW}

\date{\today}

\begin{abstract}
At cryogenic temperatures, the photoluminescence spectrum of monolayer WSe$_2$ features a number of lines related to the recombination of so-called localized excitons.
The intensity of these lines strongly decreases with increasing temperature. In order to understand the mechanism behind this phenomenon we carried out a
time-resolved experiment, which revealed a similar trend in the photoluminescence decay time. Our results identify the opening of additional nonradiative relaxation channels as a primary cause of the observed temperature quenching of the localized excitons' photoluminescence.
\end{abstract}


\maketitle

\section{Introduction}
Atomically thin semiconducting transition metal dichalcogenides (s\nobreakdash-TMDs) are nowadays a group of widely investigated two-dimensional (2D) materials that arouse a worldwide interest due to their unusual optical and electronic properties \cite{mak2010atomically,wang2012electronics}.
Monolayer s\nobreakdash-TMDs have a graphene-like 2D hexagonal crystal structure, but in contrast to graphene they possess a direct bandgap in the visible or infrared range, which immediately makes them useful for opto-electronic applications, for inastance as light emitters and photodetectors \cite{eda2013two}. Furthermore, due to absence of the lattice inversion center, their band structure has two non-equivalent valleys at the K and K' points of the Brillouin zone \cite{xiao2012coupled,cao2012valley,mak2012control,liu2015electronic}. Unlike the majority of conventional semiconductors, this so-called "valley degree of freedom" is directly coupled to specific circular polarizations of optical transitions in these materials \cite{xiao2012coupled,jones2013optical,mak2018light}. These two properties of s\nobreakdash-TMDs, \textit{i.e.} the presence of a non-zero bandgap and the valley degree of freedom directly coupled to the light, are very useful for applications in optoelectronics and its newly-opened branch focused on valley properties, which, in analogy to spintronics, is called "valleytronics". In spintronics, the storage, manipulation, and detection of information is based on the electron's spin degree of freedom \cite{johnson1985interfacial,baibich1988giant,binasch1989enhanced,Flatte2011,nature2014,Dietl2014}, while s\nobreakdash-TMDs offer an additional possibility to encode information into the K or K' valley the electron or hole resides in, or into a quantum-mechanical superposition of both valleys in the case of quantum information)\cite{wang2014, mak2012control, wu2013vapor, wang2018colloquium, koperski2017optical}. However, the information storage in s\nobreakdash-TMD-based devices still remains very challenging due to extremely short lifetime of photocreated free electron-hole pairs (excitons). 

Many interesting optical properties of s\nobreakdash-TMD materials arise to a large extent from the spin-orbit splitting of the conduction band. The magnitude of this splitting as well as the ordering of subbands are strongly material-dependent, which in connection with a much more pronounced and robust spin-orbit splitting of the valence band, results in two possible ground excitonic states: dark (optically inactive) or bright (optically active) exciton. \cite{arora2015excitonic,liu2015electronic,selig2018dark}

Tungsten diselenide (WSe$_2$) is one of s\nobreakdash-TMDs with a dark exciton ground state. A low-temperature photoluminescence (PL) spectrum of the WSe$_2$ monolayer exhibits a well-established emission pattern consisting of several relatively broad peaks related to recombination of different excitonic complexes. The highest energy peak is attributed to the optically active neutral excitonic resonance ($X$), whose decay time at low temperature is extremely short ($\sim$single ps). Fast dynamics is also characteristic of the second-in-energy feature attributed to recombination of the negatively charged exciton ($X^-$) \cite{robert2016exciton,yan2014photoluminescence,he2014tightly,you2015observation}. The PL decay times of those resonances increase with temperature to over 380 ps at about 300~K \cite{mohamed2017long}.

Apart from those short-lived free excitonic lines, the monolayer (1 ML) WSe$_2$ PL spectrum shows at lower energies a relatively broad emission band consisting of a few peaks which in literature are usually attributed to localized excitons \cite{Wang_PRB_2014,zhu2014exciton,koperski2015m,srivastava2015optically,lippert2017influence}. Importantly, compared to free excitonic resonances \cite{zhu2014exciton,mitioglu2015optical}, those states exhibit much longer lifetime, likely due to trapping effects which are believed to be involved in their formation. This makes them much more promising from the viewpoint of information storage and possibility of controlling their valley degree of freedom, which plays a crucial role in valleytronics applications \cite{smolenski2016tuning,li2019electrically,zhu2014exciton}. However, in order to fully benefit from the advantages of localized excitons, two major problems, namely their non-radiative recombination (depopulation) and decoherence, have to be overcome first. 

In contrast to neutral and charged exciton lines, the PL features of localized states in tungsten-based s\nobreakdash-TMD monolayers typically show a strong temperature dependence. At liquid helium temperature they usually dominate the PL spectrum, but almost completely disappear at a few tens of Kelvins. In general, this is related to some non-radiative carrier depopulation processes, however, the details of those processes still remain unknown. In this work we use time-integrated and time-resolved spectroscopy under pulsed excitation to reveal the mechanism responsible for the localized excitons' PL temperature dependence. Our findings shed more light on the formation and depopulation processes of localized excitonic states in atomically-thin tungsten-based s\nobreakdash-TMD materials.

\section{Results and discussion}

\begin{center}
\begin{figure}[htp]
{\includegraphics{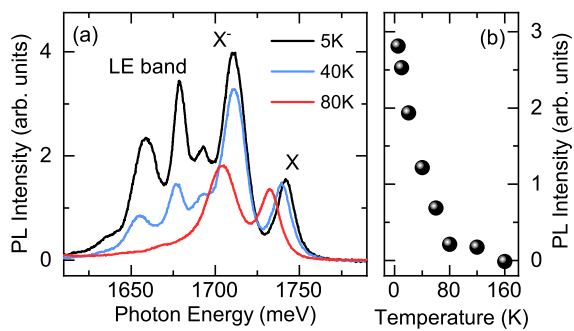}
\caption{(a) PL spectrum of WSe$_2$ monolayer measured at different temperatures ranging from 5K to 80K. In contrast to free neutral (X) and charged exciton (X$^{-}$) lines, the intensity of PL features related to localized excitons (LE) rapidly decreases at higher temperatures. (b) Integrated PL intensity of the LE band as a function of temperature.}
\label{smol}}
\end{figure}
\end{center}

\begin{center}
\begin{figure}
{\includegraphics{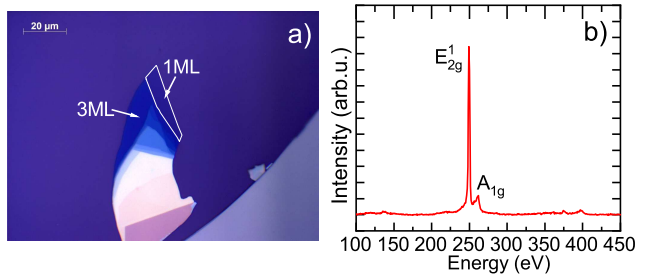}
\caption{a) Optical microscope image of one of studied WSe$_2$ flakes containing a 1 ML terrace. The monolayer's edges are marked with a white line b) Room-temperature Raman scattering spectrum of the 1 ML terrace from panel (a), obtained under 514.5 nm continuous-wave laser light excitation.}
\label{Raman}}
\end{figure}
\end{center}

WSe$_2$ flakes were prepared by mechanical exfoliation and transferred onto Si/(300~nm)SiO$_2$ substrates by means of an all-dry viscoelastic stamping technique (see Methods for details). Monolayers were then identified under an optical microscope based on their characteristic reflection contrast. The monolayer thickness of flakes selected for spectroscopic experiments was ultimately confirmed by Raman scattering measurements. Figure \ref{Raman} a) shows an optical microscopic image of one of studied monolayers, while in Fig. \ref{Raman} b) its Raman scattering spectrum is presented. The most convincing fingerprint of 1 ML thickness is the absence of a mode near 308 cm$^{-1}$, visible in all thicker WSe$_2$ films up to the bulk limit and interpreted as a combination of a shear and $E_{\text{2g}}^{1}$ modes \cite{tonndorf2013photoluminescence, zhao2013lattice}.

The experiments presented here were performed on several different WSe$_2$ flakes showing consistent results. We measured time-integrated and time-resolved, low-temperature (in the range from 5\,K to 80\,K) PL response excited non-resonantly at about 3060 meV (405 nm) with femtosecond laser pulses. Typical low-temperature spectra show a well-established pattern of emission lines with two distinct types of features (see Fig. \ref{smol}(a)). The first one consists of short-lived (a few ps) lines corresponding to free excitonic states - neutral exciton (X) at 1750 meV and charged exciton (X$^-$) at 1725 meV (at 5~K). The second type occuring in the lower-energy part of the spectrum is a long-lived ($\sim$80~ps) wide PL band of lines related to the so-called localized excitons (LEs) \cite{Wang_PRB_2014,zhu2014exciton,koperski2015m,srivastava2015optically}. The PL intensity ratio of free excitonic lines to LE peaks differs between different monolayers under study and spots on each of them, but a general pattern of the abovementioned features remains the same. With increasing temperature the linewidth of all peaks in the spectrum increases. However, in striking contrast to free excitonic lines, the PL intensity of LE lines is strongly decreasing. At about 80~K this luminescence becomes negligible compared to PL of free excitonic lines, as can be seen in Fig. \ref{smol}(a). Time- and energy-integrated PL intensity of the band related to LE states is presented in Fig. \ref{smol}(b).
Another common property to all LE states is the process leading to their population. Recent studies have shown that under non-resonant excitation, as in our experiment, the relaxation of carriers towards the localized states involves an intermediate state \cite{smolenski2016tuning} which is the same for all LEs.
Since the PL features corresponding to LEs can only be seen in systems with dark exciton ground state, such a state has been indicated as a good candidate for the intermediate state.  

\begin{center}
\begin{figure}[htp]
{\includegraphics{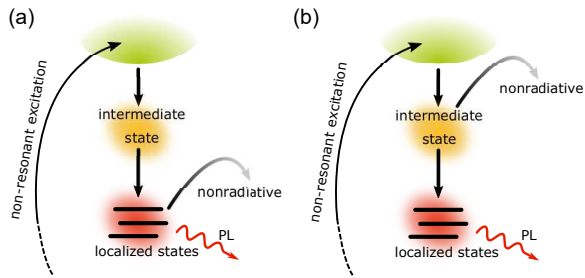}
\caption{Sketches of states and transitions contributing to the studied process of temperature quenching of the localized states' photoluminescence. Panels (a) and (b) depict two scenarios of non-radiative relaxation of intermediate and localized states, respectively.}
\label{model}}
\end{figure}
\end{center}

\begin{center}
\begin{figure}[htp]
{\includegraphics[width=1\columnwidth]{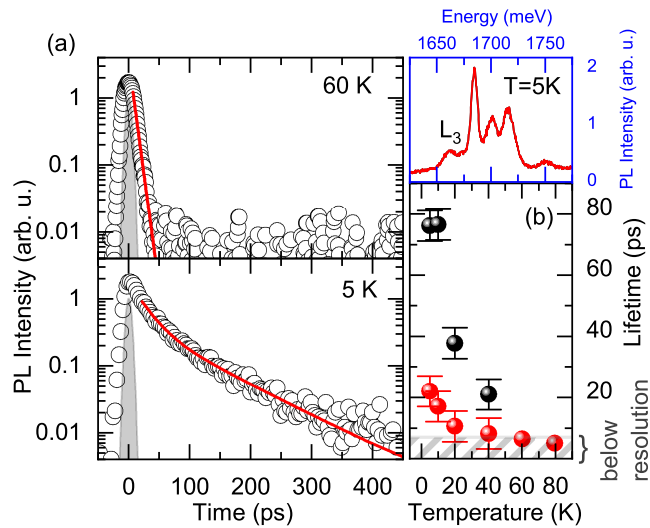}
\caption{(a) Temporal profiles of monolayer WSe$_2$ PL response (open circles) measured under pulsed non-resonant excitation at 60~K and 5~K, integrated for one LE state (L$_3$), and fitted with exponential and biexponential decays, respectively (solid red lines) (b) Temperature dependence of the L$_3$ state PL lifetime. Black and red solid circles correspond to long- and short-time components, respectively. For each LE feature the characteristic decay time of both components decreases with temperature.}\label{dec}}
\end{figure}
\end{center}

These two characteristic features of LE states: the observed decrease of PL intensity with increasing the temperature and the existence of an intermediate state in the population process rise a question about the connection between them. In general, the decrease of PL intensity is caused by the opening of additional non-radiative depopulation channels for photo-created carriers. In the situation discussed here, two different scenarios, depicted in Fig. \ref{model}, are possible: the depopulation channel can open either directly for relaxed LE states or for the intermediate state, which supplies the LE states during the relaxation process. In the former case (Fig. \ref{model}(a)), the decay time of LE luminescence features should be strongly affected by the temperature of the sample, since a hypothetical thermally-activated relaxation channel would need to be fast enough to depopulate the LE states during their lifetime. Therefore, the PL intensity should be simply inversely proportional to the PL lifetime. In the second scenario (Fig. \ref{model}(b)), only the probability of populating the LE states should decrease with increasing temperature because once populated, these states could only depopulate via the radiative recombination channel. Thus the PL lifetime should remain temperature-independent. The only temperature-related effect could in principle be observed on the rising slope of the LEs' PL temporal profile. If the intrinsic relaxation time from the intermediate state to LEs is longer than the lifetime of the intermediate state (shortened at elevated temperatures by non-radiative relaxation process), a decrease of rising time in the LEs' PL temporal profile should be observed. However, since the lifetime of intermediate state and relaxation time to LEs are very short (just a few ps) those effects are below our experimental resolution.

To determine which of the two above-mentioned depopulation channels plays a dominant role, we carried out time-resolved and time-integrated spectroscopy measurements. Figure \ref{dec} (a) shows example temporal profiles of PL signal integrated over the energy range corresponding to one of the LEs emission peaks ($L_3$) measured at 60~K and 5~K after sub-ps excitation. In the lowest temperature range, the decay is relatively slow (tens of ps) and can be well reproduced with a biexponential curve. When the temperature is changed, the characteristic times vary between 20 ps and 90 ps for the longer component and between 5\,ps and 20\,ps for the shorter component, where the higher values are observed at lower temperatures. For different monolayers under study and different LE peaks, the lifetimes may differ in value but exhibit qualitatively consistent behavior. With increasing temperature, the PL decays significantly shorten. Above 50~K distinguishing between the two components becomes impossible and the decay profiles are well reproduced with a single exponential curve. The temperature dependence of the characteristic time of both components ($\tau_1$, $\tau_2$) of the LEs' PL decay is shown in Fig. \ref{dec} (b). A significant decrease of the lifetime with the increase of temperature is observed for all LE peaks and for all studied WSe$_2$ monolayer flakes.

%As it was previously introduced there are two possible scenarios resulting in the quenching of PL from LEs. In the first case the non-radiative relaxation channel for the excitons in the intermediate state would cause the depopulation of the carriers in LE states and hence decrease of their PL intensity. This temperature effect would have no imprint on LEs PL lifetime. 

%In the second scenario we assume the inflow from the intermediate state to LEs to be temperature independent. Here we attribute the suppression of the PL from LEs to the depopulation of carriers via thermally opened non-radiative depopulation channel. This explanation is consent with our experiment - a lifetime of the LEs' PL shortens with the increase of the temperature. 

\begin{center}
\begin{figure}[htp]
{\includegraphics{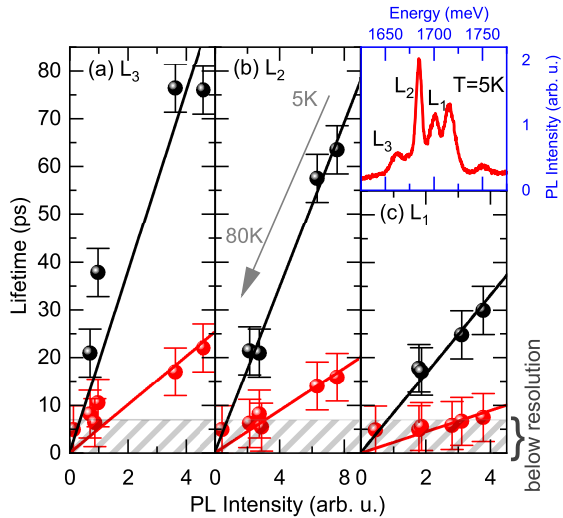}
\caption{(a-c) PL lifetime versus PL intensity for each emission line related to localized states measured at different temperatures. Shown in the inset is PL spectrum recorded at 5\,K.}
\label{AT}}
\end{figure}
\end{center}

The temperature dependence of the LEs' PL decay suggests that the decrease of the PL intensity is mainly caused by the existence of a thermally-activated depopulation channel for localized excitons. To determine if the non-radiative depopulation of the intermediate state plays any role in the PL decay, or if there are important mechanisms other than the above-mentioned ones, one needs to quantitatively compare the PL lifetime of each localized excitonic line with its intensity. Such a comparison based on data acquired on a representative monolayer at different temperatures, is shown in Fig. \ref{AT}. It clearly demonstrates that in our experiment the PL lifetime is directly proportional to time-integrated PL intensity, both measured under pulsed excitation. This linear dependence is observed for both components of the PL decay profile and for all PL lines related to localized states. Furthermore, such observation is valid for all studied monolayers.

The same, non-linear temperature dependence of the lifetime and PL intensity results in a linear dependence between the lifetime and PL intensity for both the lifetime components. The linear dependence of the PL lifetime on the PL intensity means that at all studied temperatures the same amount of carriers is injected into the LEs from the intermediate state. A part of these carriers do not recombine radiatively due to thermal opening of the non-radiative depopulation channel. Such a linear dependence leads us to the conclusion that depopulation of the intermediate state is negligible compared to dominant depopulation of localized excitonic states. This observation corresponds to the first scenario presented in Fig. \ref{model}. In the other scenario of opening the non-radiative depopulation channel from the intermediate state, less amount of carriers would be transferred to the LE states with a decrease of the PL intensity as a result, but with no change of the PL lifetime.

\section{Methods}
WSe$_2$ flakes were mechanically exfoliated from commercially available bulk crystals using a two-step, tape- and polydimethylsiloxane-based method referred to in the main text as an all-dry viscoelastic stamping technique. After the exfoliation, the flakes were indeterministically transferred onto chemically cleaned and oxygen plasma ashed Si/(300~nm)SiO$_2$ substrates. WSe$_2$ monolayers were then identified under an optical microscope based on their characteristic reflectance contrast. The monolayer thickness of flakes selected for spectroscopic study was finally confirmed with room-temperature PL and Raman scattering measurements performed with the use of 514.5 nm continuous-wave laser light exciation. During the optical experiments the sample was submerged in gaseous helium at atmospheric pressure and temperatures ranging from of 5\,K to 80\,K inside a liquid-helium-cooled cryostat. All measurements were performed under femtosecond pulsed laser excitation at 3060 meV (405 nm) from frequency-doubled Ti:Sapphire Coherent Mira Seed laser operated at a repetition rate of 76 MHz. The excitation power equal to 50 $\mu$W was chosen to stay in the linear regime and to avoid sample degradation and heating. The laser beam and PL signal were delivered to and collected from the sample via an aspheric lens mounted on a three axis piezo-electric stage. The size of the laser spot was below 2 $\mu$m. The PL spectra resolved by a monochromator were recorded with the aid of a charge-coupled-device (CCD) camera or a syncroscan Hamamatsu streak camera in the case of time-resolved measurements.

\vspace{1cm}

\begin{acknowledgments}
The work has been supported by the ATOMOPTO project (TEAM program of the Foundation for Polish Science co-financed by the EU within the ERDFund and the Nanofab facility of the Institut N\'eel, CNRS. This work was also supported by the  Polish Ministry of Science and Higher Education via a research grant "Diamentowy Grant" under decision  0149/DIA/2017/46.
 \\
 \\
\textbf{This is the version of the article as submitted by an author to Nanotechnology. IOP Publishing Ltd is not responsible for any errors or omissions in this version of the manuscript or any version derived from it.}
\end{acknowledgments}

\end{document}